\pdfoutput=1

\documentclass[11pt,twoside,a4paper,cmspaper,final,collab]{cms-tdr}
\def\svnVersion{263495}\def\cmsCernNo{2013-037}\def\cmsCernDate{\today}\def\cmsMessage{Published in Physics Letters B as \href{http://dx.doi.org/10.1016/j.physletb.2014.09.048}{\doi{10.1016/j.physletb.2014.09.048.}}}

\begin{document}\cmsNoteHeader{EXO-13-003}

\hyphenation{had-ron-i-za-tion}
\hyphenation{cal-or-i-me-ter}
\hyphenation{de-vices}

\RCS$Revision: 261156 $
\RCS$HeadURL: svn+ssh://svn.cern.ch/reps/tdr2/papers/EXO-13-003/trunk/EXO-13-003.tex $
\RCS$Id: EXO-13-003.tex 261156 2014-09-19 20:56:39Z alverson $

\newcommand{\jet}{\ensuremath{\text{jet}}\xspace}
\newcommand{\jets}{\ensuremath{\text{jets}}\xspace}
\newcommand{\cPqst}{{\ensuremath{\cmsSymbolFace{q}^\star}\xspace}}

\newlength\cmsFigWidth
\ifthenelse{\boolean{cms@external}}{\setlength\cmsFigWidth{0.95\columnwidth}}{\setlength\cmsFigWidth{0.6\textwidth}}
\ifthenelse{\boolean{cms@external}}{\providecommand{\cmsLeft}{top}}{\providecommand{\cmsLeft}{left}}
\ifthenelse{\boolean{cms@external}}{\providecommand{\cmsRight}{bottom}}{\providecommand{\cmsRight}{right}}
\ifthenelse{\boolean{cms@external}}{\providecommand{\CL}{CL\xspace}}{\providecommand{\CL}{CL\xspace}}

\cmsNoteHeader{EXO-13-003} 
\title{Search for excited quarks in the \texorpdfstring{$\gamma+\jet$}{photon+jet} final state in proton-proton collisions at \texorpdfstring{$\sqrt{s}=8\TeV$}{sqrt(s)=8 TeV}}

\date{\today}

\abstract{ A search for excited quarks decaying into the $\gamma+\jet$ final state is presented. The analysis is based on data
corresponding to an integrated luminosity of 19.7\fbinv collected by the CMS experiment in proton-proton collisions at $\sqrt{s}=8\TeV$
at the LHC. Events with photons and jets with high transverse momenta are selected and the $\gamma+\jet$ invariant mass distribution
is studied to search for a resonance peak. The 95\% confidence level upper limits on the product of cross section and branching fraction are evaluated as
a function of the excited quark mass. Limits on excited quarks are presented as a function of their mass and coupling strength; masses below 3.5\TeV
are excluded at 95\% confidence level for unit couplings to their standard model partners.
}
\hypersetup{%
pdfauthor={CMS Collaboration},%
pdftitle={Search for excited quarks in the photon+jet final state in proton-proton collisions at sqrt(s) = 8 TeV},%
pdfsubject={CMS},%
pdfkeywords={CMS, physics, photon, jet}}

\maketitle

\section{Introduction}

The standard model (SM) of strong and electroweak interactions is a theory that successfully describes a wide range of phenomena in particle physics.
Despite its immense success, the theory leaves many questions unanswered, which suggests that the SM may be an effective, low-energy approximation of a more fundamental
theory. Many proposals for physics beyond the SM are based on the assumption that quarks are composite objects. The most compelling evidence of quark substructure
would be provided by the discovery of an excited state of a quark. An excited quark ($\cPqst$) may couple to an ordinary quark and a gauge boson via gauge
interactions given by the Lagrangian~\cite{Baur:1987ga, Baur1989kv, Bhattacharya:2009xg}:

\ifthenelse{\boolean{cms@external}}{
\begin{multline}
\mathcal{L}_\text{int} = \frac{1}{2 \, \Lambda}\overline{q}^{\ast}_{R} \, \sigma^{\mu\nu}
\Big[g_{s}f_{s}\frac{\lambda_{a}}{2}G^{a}_{\mu\nu}\\
+gf\frac{\tau}{2}W_{\mu\nu}\;+\;g'f'\frac{Y}{2}B_{\mu\nu} \Big] q_{L} + \text{h.c.},
\label{eq:Lagrangian}
\end{multline}
}{
\begin{equation}
\mathcal{L}_\text{int} = \frac{1}{2 \, \Lambda}\overline{q}^{\ast}_{R}\, \sigma^{\mu\nu}
\left[g_{s}f_{s}\frac{\lambda_{a}}{2}G^{a}_{\mu\nu}\;+\;gf\frac{\tau}{2}W_{\mu\nu}\;+\;g'f'\frac{Y}{2}B_{\mu\nu} \right] q_{L} + \text{h.c.},
\label{eq:Lagrangian}
\end{equation}
}

where $\cPqst_\mathrm{R}$ is the excited quark field, $\sigma_{\mu\nu}$ is the Pauli spin matrix, $\Pq_\mathrm{L}$ is the quark field,
$G^{a}_{\mu\nu}$, $W_{\mu\nu}$ and $B_{\mu\nu}$ are the field-strength tensors of the SU(3), SU(2) and U(1) gauge fields,
$\lambda_{a}$, $\tau$, $Y$ are the corresponding gauge structure constants and $g_{s}$, $g$, $g'$ are the gauge coupling
constants. The compositeness scale, $\Lambda$, is the typical energy scale of these interactions, and $f_{s}$, $f$, $f'$ are unknown dimensionless
constants determined by the compositeness dynamics, which represent the strengths of the excited quark couplings to the SM partners and are usually
assumed to be of order unity. In proton-proton collisions, the production and decay of excited quarks, could occur via either gauge or contact
interactions~\cite{Baur1989kv}. The production of $\cPqst$ via gauge interactions would proceed through quark-gluon ($\Pq\Pg$)
annihilation. In this analysis, which assumes gauge interactions, excited quarks would then decay into a quark and a gauge boson ($\gamma, \Pg, \PW, \cPZ$)
and appear as resonances in the invariant mass distribution of the decay products. Many searches for excited quarks have been performed in various
decay channels~\cite{Beringer:1900zz, Aaron:2009iz, Abazov:2003tj,   Abe:1993sv, Abe:1997hm, Aaltonen:2008dn, Chatrchyan:2012tw, Chatrchyan:2013qha,
   ATLAS:2011ai, Aad:2013cva}, but no evidence of their existence has been found to date.

This Letter presents the first search by the CMS experiment for a resonance peak in the $\gamma+\jet$ final state. The data set used in
this study was collected in 2012 in proton-proton collisions at the CERN LHC at a center-of-mass energy of $\sqrt{s}=8\TeV$
and corresponds to an integrated luminosity of 19.7\fbinv.

Only spin-$1/2$, mass degenerate excited states of the first generation quarks, $\cPqst({=}\cPqu^\star, \cPqd^\star)$, which would be expected to be
predominantly produced in pp collisions, are considered~\cite{Baur1989kv, Bhattacharya:2009xg}. We focus on the scenario where the compositeness scale
is the same as the mass of the excited quark, \ie, $\Lambda=M_{\cPqst}$ and assume that $f_{s}, f,$ and $f'$ have the same value, denoted by $f$.

The dominant background for this search is SM $\gamma+\jet$ production. This process is an irreducible background, which is produced at leading order (LO) through quark-gluon
Compton scattering ($\Pq\Pg\to \Pq\gamma$) and quark-antiquark annihilation ($\Pq\Paq\to \Pg\gamma$). The second-largest
background is from quantum chromodynamics (QCD) dijet and multijet production, where one of the jets with high transverse momentum $\pt^{\jet}$ mimics an isolated
photon. This background falls rapidly with the photon transverse momentum $\pt^{\gamma}$ as compared to the $\gamma+\jet$ background.
The electroweak production of $\PW/\cPZ+\gamma$ would yield similar final states, but owing to their small cross section, these backgrounds are negligible.

\section{CMS detector}

The CMS experiment uses a right-handed coordinate system, with the origin at the nominal interaction point, the $x$ axis pointing to the center of the LHC, the
$y$ axis pointing up (perpendicular to the LHC plane), and the $z$ axis along the counterclockwise-beam direction. The polar angle $\theta$ is measured from the positive
$z$ axis and the azimuthal angle $\phi$ is measured in the $x$-$y$ plane. The central feature of the CMS apparatus is a superconducting solenoid of 6\unit{m}
internal diameter, providing a magnetic field of 3.8\unit{T}. Within the superconducting solenoid volume are a silicon pixel and strip tracker, a lead tungstate crystal
electromagnetic calorimeter (ECAL), and a brass/scintillator sampling hadron calorimeter (HCAL). The ECAL consists of nearly 76\,000 crystals and provides coverage in
pseudorapidity up to $\abs{\eta}<1.48$ in the barrel region and $1.48<\abs{\eta}<3.0$ in the two endcap regions, where pseudorapidity is defined as $\eta = -\ln[\tan(\theta/2)]$.
Each crystal subtends an area of 0.0174$\times$0.0174 in the $\eta$-$\phi$ plane in the barrel region. In the region $\abs{\eta}<1.74$, the HCAL cells have widths of 0.087
in pseudorapidity and 0.087 in azimuth. In the $\eta$-$\phi$ plane, and for $\abs{\eta}<1.48$, the HCAL cells map on to 5$\times$5 ECAL crystals arrays to form
calorimeter towers projecting radially outwards from close to the nominal interaction point. At larger values of $\abs{\eta}$, the size of the towers increases and
the matching ECAL arrays contain fewer crystals. A detailed description of the CMS detector can be found elsewhere~\cite{Chatrchyan:2008zzk}.

The CMS experiment uses a two-tier trigger system consisting of the first-level (L1) trigger and High Level Trigger (HLT). The L1 trigger, which is comprised of
custom electronics, reduces the readout rate from the bunch crossing frequency of approximately $20\unit{MHz}$ to below $100\unit{kHz}$. The HLT is a software-based
trigger system that makes use of information from all sub-detectors, including the tracker, to further decrease the event rate to about $400\unit{Hz}$. Only those events
passing the L1 trigger are considered by the HLT. In the HLT the photon trigger uses the same clustering algorithms as are used by the offline photon reconstruction.
Events used in this analysis passed a trigger that required at least one photon with transverse energy greater than 150\GeV. The trigger
is fully efficient for offline reconstructed photons with $\pt$ greater than 170\GeV.

\section{Event selection}

Each event is required to have at least one primary vertex reconstructed within $\abs{z}<24\unit{cm}$ from the center
of the detector and with a transverse distance less than 2\unit{cm} from the $z$-axis. The event reconstruction is performed
using a particle-flow algorithm~\cite{CMS:2009nxa, CMS:2010byl}, which reconstructs and identifies individual particles using
an optimized combination of information from all sub-detectors. Photons are identified as energy clusters in the ECAL.
These energy clusters are merged to form superclusters that are 5 crystals wide in $\eta$, centered around the most energetic crystal, and
have a variable width in $\phi$. The energy of charged hadrons is determined from a combination of the track momentum and the corresponding ECAL and HCAL
energy, corrected for the combined response function of the calorimeters. The energy of neutral hadrons is obtained from the corresponding
corrected ECAL and HCAL energies. For each event, hadronic \jets are formed from these reconstructed particles with the infrared- and collinear-safe
anti-\kt algorithm~\cite{Cacciari:2008gp}, using a distance parameter $\Delta R = 0.5$, where $\Delta R = \sqrt{(\Delta\eta)^{2} + (\Delta\phi)^{2}}$
and $\Delta\eta$ and $\Delta\phi$ are the pseudorapidity and azimuthal angle difference between the jet axis and the particle direction. Jet
energy corrections are applied to every jet to establish a uniform calorimetric response in $\eta$ and a more precise absolute response in $\pt^{\jet}$.
Jet energy scale (JES) corrections are derived from Monte Carlo (MC) simulations, and a residual correction is
derived from data~\cite{Chatrchyan:2011ds}.

Events are required to have at least one photon in the barrel region that has $\pt^{\gamma} > 170\GeV$. Photons (which can include those from $\pi^{0}$ decays
or from electron bremsstrahlung) are identified as objects associated with ECAL energy clusters not linked to the extrapolation of any charged-particle
trajectory to the ECAL. They are further required to have an ECAL shower energy profile consistent with that of a photon. The photon with the highest $\pt$ (leading)
in the event is selected as the photon candidate. The photon candidates must also satisfy the following isolation criteria:
(a) the energy deposited in the single HCAL tower closest to the supercluster position, inside a cone of $\Delta R=0.15$ centered
on the photon direction, must be less than 5\% of the energy deposited in that ECAL supercluster;
(b) the total $\pt$ of photons within a cone of $\Delta R=0.3$, excluding strips of width $\Delta\eta=0.015$ on each side of the supercluster, must be less than $0.5\GeV + 0.005\pt^{\gamma}$;
(c) the total $\pt$ of all charged hadrons within a hollow cone of $0.02<\Delta R<0.3$ about the supercluster must be less than 0.7\GeV;
(d) the total $\pt$ of all neutral hadrons within a cone of $\Delta R=0.3$ must be less than $0.4\GeV + 0.04\pt^{\gamma}$.
These isolation variables are corrected for the presence of additional reconstructed vertices associated with extra interactions in the same bunch
crossing (pileup) by subtracting the average energy calculated from the typical energy density in the event, as computed using the \textsc{FastJet}
package~\cite{Cacciari2011ma}. The signal efficiency is found to be ${\sim}70\%$ for the photon identification and isolation selection criteria.
Anomalous calorimeter signals~\cite{Petyt:2012upa}, caused by isolated large noise in the detector could be reconstructed as photon candidates.
A selection is therefore applied on the shower shape variables to largely remove such photon candidates from the event. In addition, to reduce
the anomalous calorimeter noise signals~\cite{reeewf}, the ECAL crystals with energy greater than 1\GeV are required to be within a 5\unit{ns}
window relative to the supercluster time.

The leading $\jet$ separated from the photon candidate by $\Delta R>0.5$ and satisfying particle flow based $\jet$
identification criteria~\cite{CMS:2010xta} is selected as the $\jet$ candidate. The $\jet$ identification criteria include requirements on
the number of constituents and on the fraction of the $\jet$ energy held by each constituent type. The $\jet$ candidate is required to be within
the pseudorapidity region $|\eta^{\jet}|<3.0$ and must have a transverse momentum $\pt^{\jet}>170\GeV$. The invariant mass of $\GAMJET$ is
calculated using the leading photon and jet candidates and is given by $M_{\gamma,\jet} = \sqrt{ (E^{\gamma} + E^{\jet})^2 - (\vec{p}^{\gamma} + \vec{p}^{\jet})^2 }$,
where $E$ and $\vec{p}$ denote the energy and momentum, respectively, of the photon and of the jet.

The production of excited quarks via the expected s-channel process would result in an isotropic distribution of final-state objects.
All backgrounds are produced predominantly through t-channel processes and have
an angular distribution that is strongly peaked in the forward or backward direction. Therefore, to reduce these backgrounds while retaining high
signal acceptance, the leading photon and $\jet$ candidates are required to satisfy $\abs{\Delta\eta(\gamma,\jet) } < 2.0$. To ensure the back-to-back
topology expected in a two body final state, $\abs{\Delta\phi(\gamma,\jet)} > 1.5$ is required between the photon and $\jet$ candidates. The
above-mentioned thresholds for $\abs{\Delta\eta}$, $\abs{\Delta\phi}$, and $\abs{\eta^{\jet}}$ selection were chosen to optimize the search sensitivity.
A selection on the mass, $M_{\gamma,\jet}> 560\GeV$, is
applied to avoid the kinematical turn-on region associated with the various selection requirements.

\section{Resonance shape and background fit}

\begin{figure}[h!]
\centering
\includegraphics[width=\cmsFigWidth]{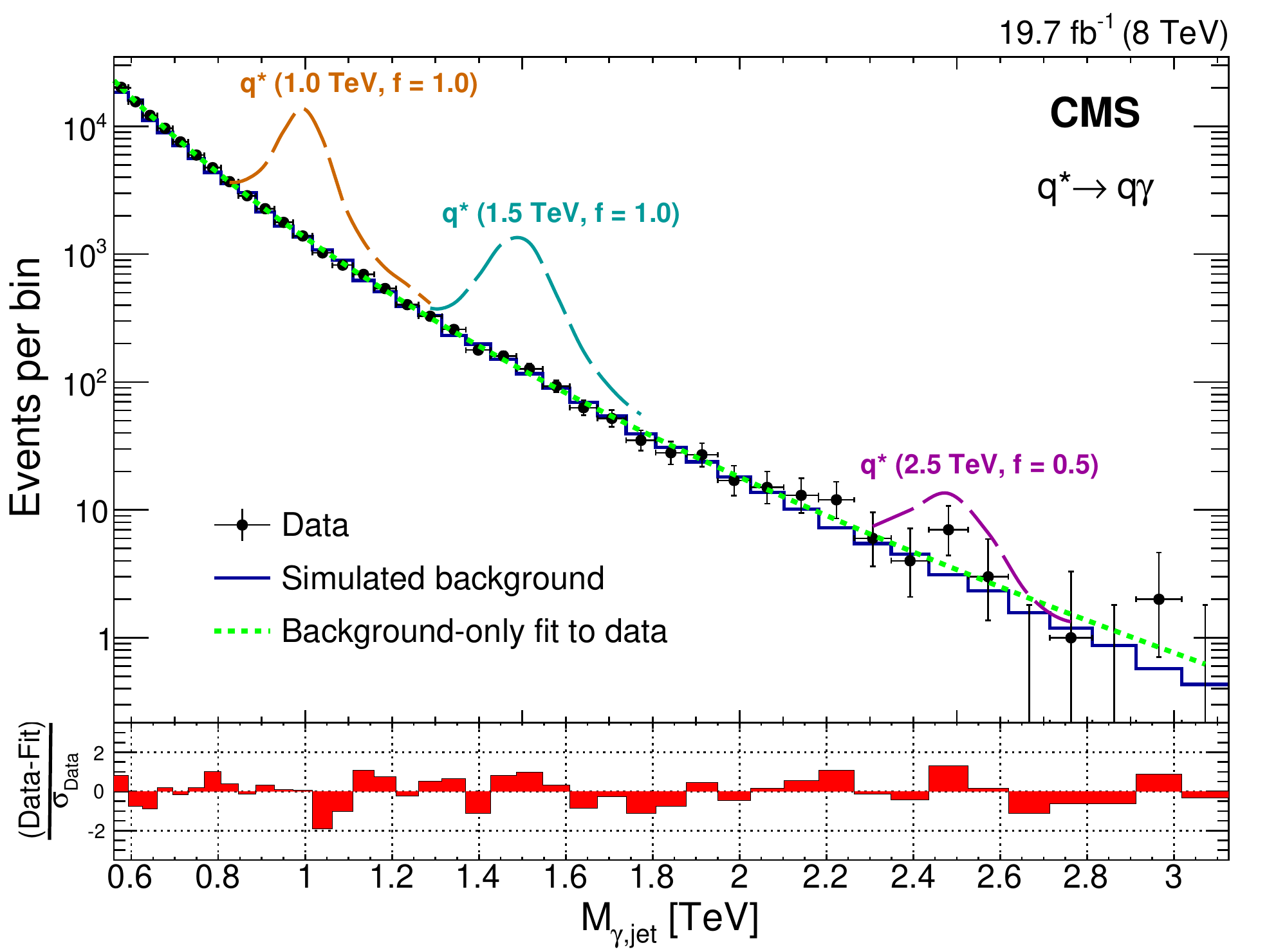}
\caption{The $\gamma+\jet$ invariant mass distribution in data (points) and MC prediction (histogram) after full selection. The horizontal bar on each data point denotes
   the bin width. The asymmetric error bars indicate central confidence intervals appropriate for Poisson-distributed data and are obtained from the Neyman construction
      as described in~\cite{Garwood}. The result of the fit to the data using the background parameterization of Eq.~\eqref{eq:bkgfit} is shown with the dotted green curve.
      The bin-by-bin fit residuals, (Data-Fit) divided by the statistical uncertainty in the data ($\sigma_{\text{Data}}$), are shown at the bottom. The bin-widths
      used reflect the expected mass resolution. Mass distributions for three sample signal values (mass and couplings) are also shown.
}
\label{fig:massFit}
\end{figure}

The invariant mass distributions of the $\gamma+\jet$ events in the collected data and for simulated events, after applying all the selections,
are shown in Fig.~\ref{fig:massFit}. The $\gamma+\jet$ and dijet MC predictions are generated using \PYTHIA 6.426~\cite{Sjostrand:2006za},
based on a LO calculation, while the electroweak backgrounds are taken from the \MADGRAPH~\cite{Alwall:2014hca} event
generator. The underlying event tune Z2$^\star$~\cite{Chatrchyan:2011id,Field:2010bc} and the CTEQ6L1~\cite{Pumplin:2002vw,Martin:2009iq} parton distribution
functions (PDFs) are used. The generated events are processed with a full detector simulation based on  \GEANTfour~\cite{Agostinelli:2002hh} and
the same event reconstruction package as used for data. The MC prediction is normalized to the integrated luminosity of the data sample. A K-factor
of 1.3~\cite{Ellis:1992en,Giele:1993dj} is used to scale the \PYTHIA $\gamma+\jet$ and dijet predictions to account for the
next-to-leading-order contributions. A correction factor of 0.95 is applied to account for an observed difference in the efficiencies
of the photon identification requirements in data and MC simulation. After the full selection is applied, it is estimated that SM $\gamma+\jet$ production
accounts for 80.5\% of the total background, 18.5\% comes from dijets, while electroweak background contributes 1.0\%.

The background-only MC simulation, while not used to obtain the analysis results, is seen in Fig.~\ref{fig:massFit} to describe the data well, both in shape and yield. The mass distributions of both simulations and data, shown in Fig.~\ref{fig:massFit}, are plotted in bins of width equal to the expected mass resolution, which varies from 4.5\% at 1\TeV to 3\% at 3\TeV. The highest mass event observed in data is at 2.9\TeV.

The expected signal from excited quarks produced via $\Pq\Pg$ fusion is simulated using the LO calculation available in \PYTHIA 6.426.
The signal mass distributions for three $\cPqst$ values after full reconstruction and selection are shown in Fig.~\ref{fig:massFit}. The same
underlying event tunes of Z2$^\star$ and CTEQ6L1 PDF are used as for the background MC events. Two different coupling scenarios, $f=1.0$ and $0.5$ are
considered. The cross section scales as $f^{2}$ and the natural width of the resonance peak can be approximated as ${\sim}0.04 f^{2} M_{\cPqst}$, although the
observed width is dominated by the experimental $\gamma+\jet$ mass resolution and is therefore independent of $f$ for $f\leq1$.

As described in the following section, the analysis compares the observed data with a background determined from an analytic fit to the data plus the possible
presence of a signal. The modeling of the SM photon and jet background mass distribution is based on the parameterization:

\begin{equation}
\frac{\rd\sigma}{\rd{}m} = \frac{P_{0}(1-m/\sqrt{s})^{P_{1}}}{(m/\sqrt{s})^{P_{2}+P_{3}\ln(m/\sqrt{s})}},
\label{eq:bkgfit}
\end{equation}

where $\sqrt{s}=8\TeV$, and $P_{0}$, $P_{1}$, $P_{2}$, and $P_{3}$ are the four parameters used to describe the background. This functional form
has been widely used in similar previous searches~\cite{Aaltonen:2008dn,Chatrchyan:2013qha, ATLAS:2011ai, Aad:2013cva}. It is motivated by the functional form of QCD,
with a term in the numerator that mimics the mass dependence of parton distributions, and a term in the denominator that mimics the mass dependence of the QCD matrix
element. The parameterization in Eq.~\eqref{eq:bkgfit} gives a good description of both the simulated background distribution and the observed data, as may be
seen in Fig.~\ref{fig:massFit}. The resulting fit to the data after final selection, shown in Fig.~\ref{fig:massFit}, has a $\chi^{2}$ of 20.57 for 34 degrees
of freedom. The residual difference between data and fit for each mass bin is shown at the bottom of Fig.~\ref{fig:massFit}. No significant differences between
the data and the background-only fit are observed.

\section{Results}

A Bayesian formalism~\cite{Beringer:1900zz} using a binned likelihood with a uniform prior for the signal cross section is
used for estimating the upper limit on the cross section. The data are fit to the background function given by Eq.~\eqref{eq:bkgfit} plus the signal line shape from MC simulation,
with the signal cross section treated as a free parameter.  The resulting fit function with the signal cross section set to zero is used as the background hypothesis.
Log-normal prior distribution functions are used to model the systematic uncertainties which are treated as nuisance parameters in the limit setting procedure.
For each resonance mass ranging from 0.7\TeV to 4.4\TeV in steps of 0.1\TeV, the posterior probability density is calculated as a function of signal cross section.
Finally, the 95\% confidence level (\CL) upper limit is calculated from the posterior probability density at each mass point.

The accounted systematic uncertainties include $\jet$ energy resolution (JER) (10\%)~\cite{Chatrchyan:2011ds}, photon energy resolution
(PER) (0.5\%)~\cite{Chatrchyan:2013dga}, jet energy scale (JES) ($1.0-1.4\%$)~\cite{Chatrchyan:2011ds}, photon energy scale (PES)
(1.5\%)~\cite{Chatrchyan:2013dga, Chatrchyan:2013fya}, and uncertainty in the integrated luminosity (2.6\%)~\cite{CMS:2013jza}.
The systematic uncertainty associated with JER and PER translate into a 5\% relative uncertainty in the mass resolution, which is propagated
into the result by increasing and decreasing the width of the reconstructed mass shape of signal. The effects of JES and PES uncertainties
are estimated to be 0.5--0.7\% (as a function of $\gamma+\jet$ mass) and 0.7\%, respectively. These uncertainties are accounted for
by shifting the reconstructed signal mass by 1\%. The uncertainty in the integrated luminosity is included to account for the uncertainty in
the normalization of the signal. A systematic uncertainty of 4\% in the acceptance $\times$ efficiency ($A\times\epsilon$) derived from the uncertainties
in the measurement of correction factors is also included. A signal uncertainty of 0.3\% is estimated from the pileup modeling in simulation.
Theoretical uncertainties are also considered for the signal samples and include uncertainties based on differences stemming from the choice of PDF and the
factorization and renormalization scales. The systematic uncertainty from the choice of PDF for different signal resonance masses is estimated according to the PDF4LHC
recommendations~\cite{Botje:2011sn,Alekhin:2011sk,Ball:2012cx}. The factorization and renormalization scales are varied by factors of 0.5 and 2.0 and the
variation of the signal cross section for different resonance masses is evaluated. The uncertainties in the signal acceptance based on the choice of PDF and in the
cross section from variations in scales are found to be about 0.5\% and 4\%, respectively.

Other sources of systematic uncertainty include the description of initial-state radiation (ISR) and final-state radiation (FSR),
which could potentially affect the shape of the resonant peak. The effect of ISR is small and mostly contained in the low-mass tail, but
FSR could affect the mean and the width of a resonance significantly. The effect of FSR uncertainties depends on the choice of its scale~\cite{Sjostrand:2006za},
and is estimated by varying this scale by factors of 0.5 and 2.0. The change in the mean is found to be within $\pm 0.5\%$ for both low- and high-mass
signals. The change in the width is found to be 7\% for 1\TeV and 4\% for 3\TeV mass signals.

The systematic uncertainties in JER, PER, JES, PES, FSR, in the correction factors, and in the integrated luminosity are used in the limit setting procedure
as nuisance parameters and affect only the signal. The effect on the signal shape of systematic uncertainty associated with the pileup correction is negligible.
The statistical uncertainty in the fit prediction is estimated to be 1\% at 1\TeV and 30\% at 3\TeV. This uncertainty is estimated by
interpreting the number of observed events in each bin as the mean of a Poisson distribution, which is randomly sampled to generate new pseudo-data. The pseudo-data are
fit using the parameterization given in Eq.~\eqref{eq:bkgfit}. This procedure is repeated many times and the fit uncertainty is taken as the maximal deviation
observed from the nominal fit. For the background shape uncertainty, the background parameters are marginalized with a flat prior. The effect of these systematic
uncertainties is small in the region of high masses, relevant to the estimation of the lower bound on $M_{\cPqst}$. Thus the extracted limits are robust.

\begin{table*}[h!]
\centering
\topcaption{The expected and observed 95\% \CL upper limits on $\sigma\times\mathcal{B}$ for the production of excited quarks in the $\GAMJET$ final state, assuming a coupling strength $f=1.0$.}
\begin{tabular}{rccrcc}
\hline
Mass    & \multicolumn{2}{c}{Upper limit (fb)} &  Mass  & \multicolumn{2}{c}{Upper limit (fb)} \\
(\TeVns)   & Expected & Observed & (\TeVns) & Expected & Observed \\
\hline
0.7  &  76.4 &  93.2 & 2.6 &  1.11 &  1.22 \\
0.8  &  49.8 &  59.0 & 2.7 &  0.96 &  0.75 \\
0.9  &  34.3 &  24.9 & 2.8 &  0.84 &  0.60 \\
1.0 &  25.2 &  13.5 & 2.9 &  0.76 &  0.58 \\
1.1 &  17.6 &  13.6 & 3.0 &  0.72 &  0.61 \\
1.2 &  13.8 &  17.9 & 3.1 &  0.70 &  0.51 \\
1.3 &  11.1 &  14.1 & 3.2 &  0.62 &  0.43 \\
1.4 &   8.57 &  10.6 & 3.3 &  0.57 &  0.39 \\
1.5 &   6.52 &  10.5 & 3.4 &  0.55 &  0.34 \\
1.6 &   5.59 &   7.15 & 3.5 &  0.54 &  0.35 \\
1.7 &   4.71 &   3.98 & 3.6 &  0.51 &  0.35 \\
1.8 &   3.87 &   2.72 & 3.7 &  0.48 &  0.34 \\
1.9 &   3.05 &   2.82 & 3.8 &  0.45 &  0.33 \\
2.0 &   2.60 &   2.72 & 3.9 &  0.43 &  0.32 \\
2.1 &   2.18 &   2.84 & 4.0 &  0.45 &  0.34 \\
2.2 &   1.80 &   2.79 & 4.1 &  0.44 &  0.34 \\
2.3 &   1.56 &   2.29 & 4.2 &  0.44 &  0.34 \\
2.4 &   1.45 &   1.86 & 4.3 &  0.42 &  0.34 \\
2.5 &   1.32 &   1.66 & 4.4 &  0.42 &  0.34 \\
\hline
\end{tabular}
   \label{Table:ObsLimits}
\end{table*}

\begin{figure}[h!]
\centering
\includegraphics[width=\cmsFigWidth]{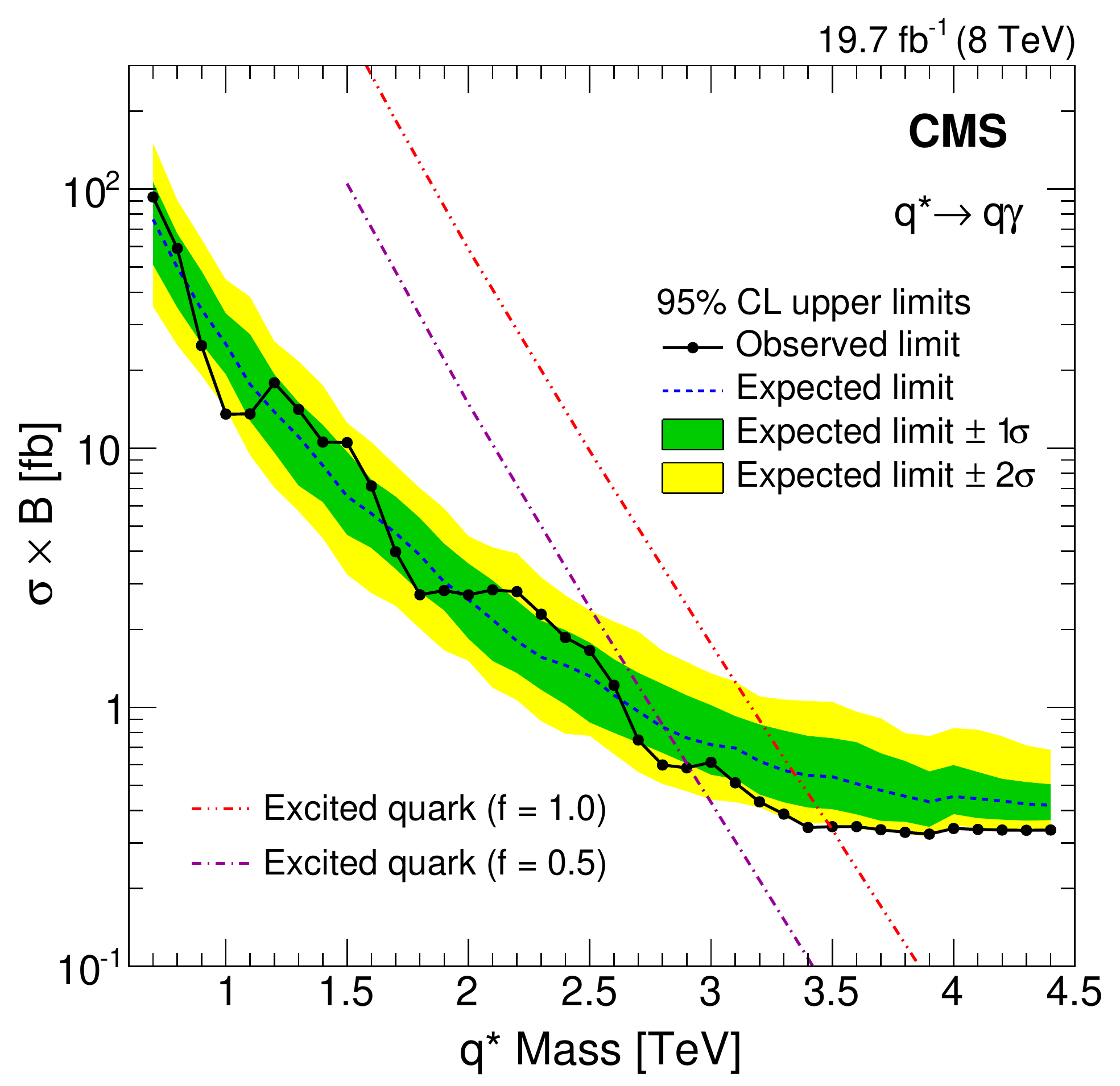}
\caption{The expected and observed 95\% \CL upper limits on $\sigma\times\mathcal{B}$ for the production of excited quarks in the $\gamma+\jet$ final state. The limits
are also compared with theoretical predictions for excited quark production, shown for two values of the coupling strengths. The uncertainty in the expected limits at the $1\sigma$
and $2\sigma$ levels are shown as shaded bands.}
\label{fig:fullhalfSigmaBR}
\end{figure}

The 95\% \CL upper limit on $\sigma\times\mathcal{B}$ as a function of $M_\cPqst$ is listed in Table~\ref{Table:ObsLimits} and
shown in Fig.~\ref{fig:fullhalfSigmaBR}. For signal, $A \times \epsilon$ is found to range from 54 to 57\% for $\cPqst$ masses from
1 to 4\TeV. The observed limits are found to be consistent with those expected in the absence of a signal. These limits are evaluated up to a $\cPqst$
mass of 4.4\TeV, since at higher values of $M_{\cPqst}$, off-shell production dominates, thus reducing the sensitivity of the search. This behavior
agrees with that reported in~\cite{Aad:2013cva}.

\begin{figure}[h!]
\centering
\includegraphics[width=\cmsFigWidth]{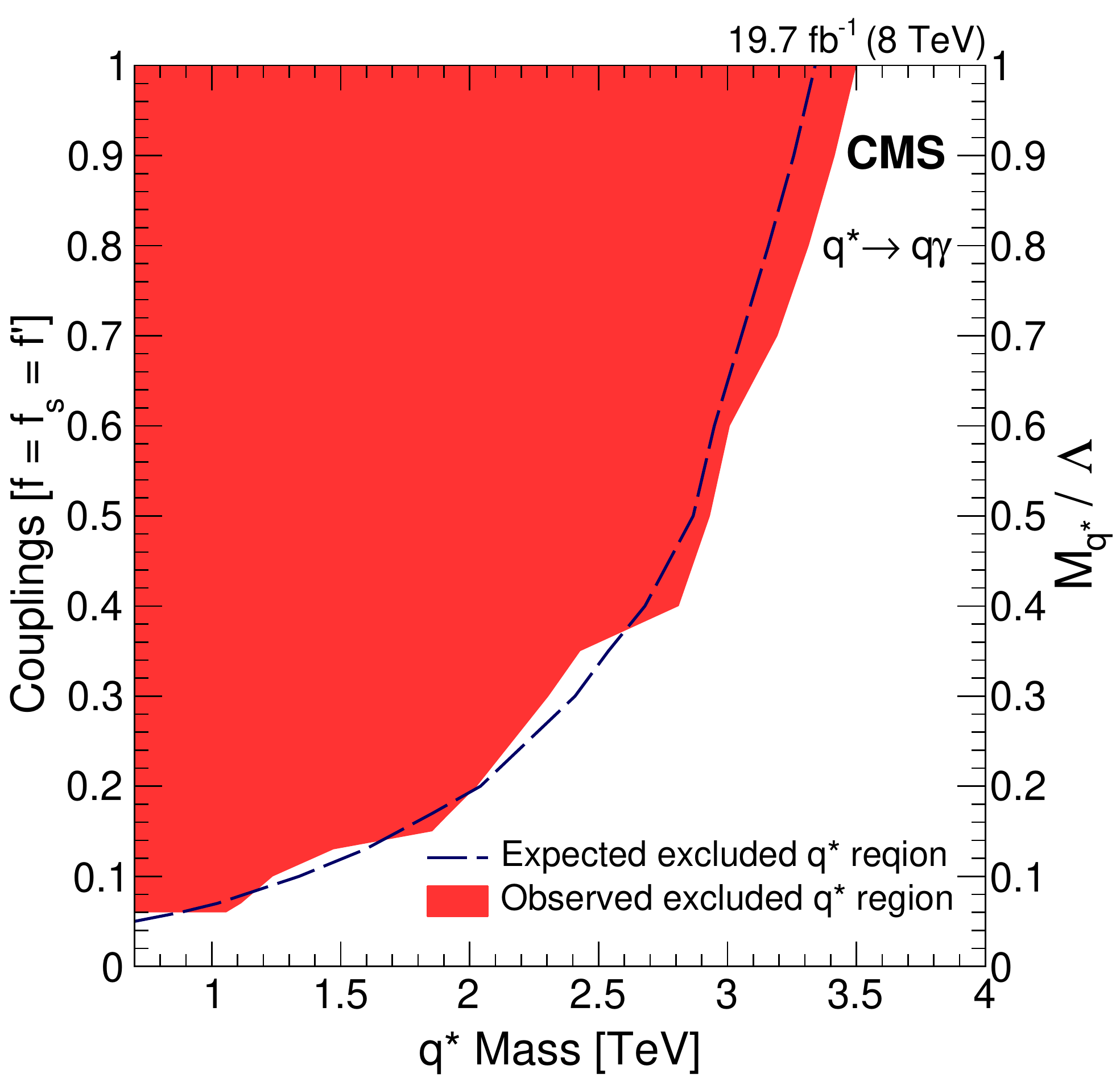}
\caption{The observed (red filled) and expected (dashed line) excluded regions at 95\% \CL as a function of excited quark mass and the coupling strength for $\Lambda = M_\cPqst$ (left axis) or the mass divided by compositeness scale for coupling strength of $f=1$ (right axis). }
\label{fig:CouplingMass}
\end{figure}

The observed limits are compared to the LO theoretical predictions, shown in Fig.~\ref{fig:fullhalfSigmaBR} for $f=1.0$ and 0.5,
to estimate the lower mass bounds on excited quarks. A lower bound of 3.5 (2.9)\TeV is obtained for $f=1.0$ (0.5). The
corresponding expected mass limits are 3.3 (2.8)\TeV. If we take into account the effect of the theoretical uncertainty due to variation in
factorization and renormalization scales on the signal cross section, then the observed limit on $M_{\cPqst}$ changes
by $\pm$0.2\%. The dependence of the $\sigma\times\mathcal{B}$ upper limit on $f$ is found to be negligible for $f\leq1$ since the observed
resonance width is dominated by the experimental resolution. Using the theoretical predictions for different coupling strengths from 0.1 to 1.0 and
observed limits, a mass region as a function of coupling strength is excluded, as shown in Fig.~\ref{fig:CouplingMass}.

The result shown in Fig.~\ref{fig:CouplingMass} may also be interpreted to be presenting limits on the excited quark mass as a function of compositeness
scale $\Lambda$, if the conventional assumption $\Lambda=M_{\cPqst}$ is relaxed. This is because variations in $f$ and in $M_{\cPqst}/\Lambda$
have the same effect on the $\cPqst$ cross section. For example, from Fig.~\ref{fig:CouplingMass}
if we assume $\Lambda=10M_{\cPqst}$ and SM couplings, then we exclude excited quarks with mass $0.7< M_{\cPqst}<1.2\TeV$.

\section{Summary}

A search for excited quarks in the $\gamma+\jet$ final state has been presented. The proton-proton collision data set at $\sqrt{s}= 8\TeV$ corresponds
to an integrated luminosity of 19.7\fbinv. The data are found to be consistent with the predictions of the standard model and upper limits are placed
on $\sigma\times\mathcal{B}$ for $\cPqst$ production in the $\gamma+\jet$ final state.

Comparing these limits with the theoretical predictions, excited quarks with masses in the range $0.7<M_{\cPqst}<3.5\TeV$ are excluded
at 95\% confidence level under the standard assumption $f=1.0$. These results are similar to those from a search in the $\gamma+\jet$
final state~\cite{Aad:2013cva} by the ATLAS experiment and may be compared with those from a search for excited quarks in the dijet final state
at CMS, which set a lower bound on $M_{\cPqst}$ of 3.19\TeV with 4.0\fbinv of data~\cite{Chatrchyan:2013qha}.

For the first time at the LHC, the sensitivity of the search has also been investigated for coupling strengths less
than unity, as shown in Fig.~\ref{fig:CouplingMass}. Excited quarks with masses in the range $0.7< M_{\cPqst}<2.9\TeV$ are excluded
for $f=0.5$. Furthermore, excited quark masses in the range $0.7< M_{\cPqst}<1.0\TeV$ are excluded for couplings as low as $f=0.06$.

\section*{Acknowledgements}
We thank Debajyoti Choudhury for inspiring this work and continuing to provide theoretical insight throughout its course.
We congratulate our colleagues in the CERN accelerator departments for the excellent performance of the LHC and thank the technical and administrative staffs at CERN and at other CMS institutes for their contributions to the success of the CMS effort. In addition, we gratefully acknowledge the computing centres and personnel of the Worldwide LHC Computing Grid for delivering so effectively the computing infrastructure essential to our analyses. Finally, we acknowledge the enduring support for the construction and operation of the LHC and the CMS detector provided by the following funding agencies: BMWFW and FWF (Austria); FNRS and FWO (Belgium); CNPq, CAPES, FAPERJ, and FAPESP (Brazil); MES (Bulgaria); CERN; CAS, MoST, and NSFC (China); COLCIENCIAS (Colombia); MSES and CSF (Croatia); RPF (Cyprus); MoER, ERC IUT and ERDF (Estonia); Academy of Finland, MEC, and HIP (Finland); CEA and CNRS/IN2P3 (France); BMBF, DFG, and HGF (Germany); GSRT (Greece); OTKA and NIH (Hungary); DAE and DST (India); IPM (Iran); SFI (Ireland); INFN (Italy); NRF and WCU (Republic of Korea); LAS (Lithuania); MOE and UM (Malaysia); CINVESTAV, CONACYT, SEP, and UASLP-FAI (Mexico); MBIE (New Zealand); PAEC (Pakistan); MSHE and NSC (Poland); FCT (Portugal); JINR (Dubna); MON, RosAtom, RAS and RFBR (Russia); MESTD (Serbia); SEIDI and CPAN (Spain); Swiss Funding Agencies (Switzerland); MST (Taipei); ThEPCenter, IPST, STAR and NSTDA (Thailand); TUBITAK and TAEK (Turkey); NASU and SFFR (Ukraine); STFC (United Kingdom); DOE and NSF (USA).
\ifthenelse{\boolean{cms@external}}{\vspace*{4ex}}{}
\bibliography{auto_generated}
\cleardoublepage \appendix\section{The CMS Collaboration \label{app:collab}}\begin{sloppypar}\hyphenpenalty=5000\widowpenalty=500\clubpenalty=5000\textbf{Yerevan Physics Institute,  Yerevan,  Armenia}\\*[0pt]
V.~Khachatryan, A.M.~Sirunyan, A.~Tumasyan
\vskip\cmsinstskip
\textbf{Institut f\"{u}r Hochenergiephysik der OeAW,  Wien,  Austria}\\*[0pt]
W.~Adam, T.~Bergauer, M.~Dragicevic, J.~Er\"{o}, C.~Fabjan\cmsAuthorMark{1}, M.~Friedl, R.~Fr\"{u}hwirth\cmsAuthorMark{1}, V.M.~Ghete, C.~Hartl, N.~H\"{o}rmann, J.~Hrubec, M.~Jeitler\cmsAuthorMark{1}, W.~Kiesenhofer, V.~Kn\"{u}nz, M.~Krammer\cmsAuthorMark{1}, I.~Kr\"{a}tschmer, D.~Liko, I.~Mikulec, D.~Rabady\cmsAuthorMark{2}, B.~Rahbaran, H.~Rohringer, R.~Sch\"{o}fbeck, J.~Strauss, A.~Taurok, W.~Treberer-Treberspurg, W.~Waltenberger, C.-E.~Wulz\cmsAuthorMark{1}
% ...
\vskip\cmsinstskip
% ...
1:~~Also at Vienna University of Technology, Vienna, Austria\\
2:~~Also at CERN, European Organization for Nuclear Research, Geneva, Switzerland\\
% ...

\end{sloppypar}
\end{document}